# Large magnetic entropy change near room temperature in antipervoskite SnCMn$_3$

B. S. Wang,[1] P. Tong,[1,2(a)] Y. P. Sun,[1(b)] X. Luo,[1] X. B. Zhu,[1] G. Li,[1] X. D. Zhu,[1] S. B. Zhang,[1] Z. R. Yang,[1] W. H. Song[1] and J. M. Dai[1]

[1] *Key Laboratory of Materials Physics, Institute of Solid State Physics, and High Magnetic Field Laboratory, Chinese Academy of Sciences, Hefei 230031, People's Republic of China*

[2] *Department of Physics, University of Virginia, Virginia 22904, USA*

**Abstract**- We report the observation of large magnetocaloric effect near room temperature in antipervoskite SnCMn$_3$. The maximal magnetic entropy change at the first-order ferrimagnetic-paramagnetic transition temperature ($T_C \sim 279$ K) is about 80.69mJ/cm$^3$ K and 133mJ/cm$^3$ K under the magnetic field of 20 kOe and 48 kOe, respectively. These values are close to those of typical magnetocaloric materials. The large magnetocaloric effect is associated with the sharp change of lattice, resistivity and magnetization in the vicinity of $T_C$. Through the measurements of Seebeck coefficient and normal Hall effect, the title system is found to undergo a reconstruction of electronic structure at $T_C$. Considering its low-cost and innocuous raw materials, Mn-based antiperovskite compounds are suggested to be appropriate for pursuing new materials with larger magnetocaloric effect.

*Corresponding author. Tel: +86-551-559-2757; Fax: +86-551-559-1434.

[(a)]E-mail address: tongpeng@issp.ac.cn (P. Tong),

[(b)] ypsun@issp.ac.cn (Y. P. Sun)

Materials with large magnetocaloric effect (MCE) have attracted much attention in recent years due to their potential application in magnetic refrigeration, which provides a promising alternative to conventional vapor-cycle one [1-3]. The research has been mainly focused on the materials with expensive rare-earth elements, sometimes with poisonous elements (e.g., As). Therefore, practically it is of great significance to explore new systems with large MCE, but with innocuous and low-cost raw materials.

The antiperovskite intermetallic compounds $AXM_3$ (A, main group elements; X, carbon, boron or Nitrogen; M, transition metal) have displayed lots of interesting properties, such as superconductivity [4, 5], giant magnetoresistance [6], large negative MCE [7, 8], giant negative thermal expansion (NTE) [9, 10], etc. Up to present, $GaCMn_3$ is the only antiperovskite compound that has been reported to exhibit MCE. In $GaCMn_3$ there exists an abrupt first-order antiferromagnetic (AFM)-ferromagnetic (FM) transition at ~165 K, which produces a large entropy change $-\Delta S_M$ (15 J/kg K under 20 kOe) with a plateau-like temperature dependence [7, 8]. In carbon-deficient sample $GaC_{0.78}Mn_3$, the AFM state observed in stoichiometric sample is collapsed. Instead, the ground state is FM with a transition temperature of 295 K, at which the value of $-\Delta S_M$ is 3.7 J/kg K under 50 kOe [11]. For $GaCMn_{3-x}Co_x$, the substitution of Co for Mn sites decreases the AFM-FM transition temperature without significant loss of $-\Delta S_M$, while broadens the plateau-like temperature dependence [12].

In this paper, we report large room-temperature MCE in another antiperovskite compound $SnCMn_3$, corresponding to a ferrimagnetic (FIM)-paramagnetic (PM) transition ($T_C \sim 279$ K). The large MCE is related with sharp change of magnetization at $T_C$, which a reconstruction of electronic structure can account for. Along with the previous reports on Mn-based antiperovskite materials, we suggest this type of compounds could be an alternative for searching new MCE materials at various temperatures.

Polycrystalline sample of $SnCMn_3$ was prepared from powders of Sn (4N), Graphite (3N) and Mn (4N). The starting materials were mixed in the desired mole ratio, sealed in evacuated quartz tubes and then heated at 1073K for two days, slowly increased to 1150K at 1.5K/min, and then annealed at 1150K for five days. After the tube was quenched to room temperature, the products were pulverized, grounded, and pressed into pellets which were annealed again under the same condition to obtain the homogeneous samples. X-ray powder diffraction (XRD) pattern was collected using a Philips X' pert PRO x-ray diffractometer with Cu $K\alpha$ radiation at room temperature. The XRD data was analyzed by using the standard Rietveld technique. The magnetic measurements were performed on a Quantum Design superconducting quantum interference device (SQUID) magnetometer (1.8 K ≤ $T$ ≤ 400 K, 0 ≤ $H$ ≤ 50kOe). The Seebeck coefficient and Hall effect were measured in a commercial Quantum Design Physical Property Measurement System (PPMS) (1.8K ≤ $T$ ≤ 800K, 0 ≤ $H$ ≤ 90kOe).

Fig. 1 shows the typical refinement of XRD pattern for the $SnCMn_3$ sample. All diffraction peaks can be indexed by the antiperovskite structure (space group, $Pm3m$) and no detectable secondary phases can be found. The refined lattice parameter $a$ (3.981 Å) is slightly less than the calculated one (3.989 Å) [13].

Magnetization as a function of temperature measured $M$-$T$ at a field of 100 Oe is shown in fig.2. A clear magnetic transition can be found at $T_C \sim 279$ K, which is determined as the reflection point of derivative of $M$-$T$ curve. This value of $T_C$ is roughly in accordance with previously reported one [14]. The neutron diffraction measurement has revealed that the ground state of $SnCMn_3$ is a noncollinear FIM one, which consists of both AFM and FM sublattices [15, 16]. Thus the magnetic transition we observed at $T_C$ should be a FIM-PM one. As shown in fig. 2, a weak thermal hysteresis (~2.5 K) between warming and cooling FC curves can be found near $T_C$, indicating a first-order transition. Generally, the first-order phase transition is associated with a structural transformation. As for $SnCMn_3$, with increasing temperature the unit cell volume undergoes a discontinuous contraction of ~0.1% at $T_C$, without changing the crystal structure [14]. As shown on the right hand of fig.2, the inverse magnetic susceptibility $1/\chi(T)$ above $T_C$ roughly agrees with the Curie-Weiss law. However, the $1/\chi(T)$ at elevated temperatures does not obey this law as reported by other authors with unclear reason [16].

Fig. 3 displays the temperature dependent resistivity $\rho(T)$ measured at zero field. At low temperatures, $\rho(T)$ shows a metallic behavior. With increasing temperature, $\rho(T)$ presents a sudden reduction at $T_C$. As temperature increases further, $\rho(T)$ turns back to the metallic behavior with a smaller $d\rho(T)/dT$ ratio than at low temperatures. As shown in the inset of fig. 3, an obvious thermal hysteresis between warming and cooling modes can be seen, suggesting a first-order transition near the $T_C$, which is consistent with the result of fig. 2.

Fig.4 presents the magnetization isotherms *M-H* of SnCMn$_3$ at selected temperatures near $T_C$. These *M-H* curves were measured after the sample was cooled down from 300 K to each measurement temperature. A metamagnetic transition induced by external fields can be observed clearly in the *M-H* curves around $T_C$. At several selected temperatures near $T_C$, the isotherms in the increasing field/decreasing field cycles are also measured and there are small hystereses in these cycles. Such a small hystereses of *M(H)* is significant for the practical application of magnetic refrigeration [17, 18].

Based on the *M-H* curves, the magnetic entropy change can be evaluated using the Max-well equation [2],

$$\Delta S_M(T,H) = S_M(T,H) - S_M(T,0) = \int_0^H \left(\frac{\partial S}{\partial M}\right)_T dH = \int_0^H \left(\frac{\partial M}{\partial T}\right)_H dH. \quad (1)$$

In the case of magnetization measurements at small discrete field and temperature intervals, $\Delta S_M$ can be approximated as [2],

$$\left|\Delta S_M \left(\frac{T_i + T_{i+1}}{2}\right)\right| = \sum \left[\frac{(M_i - M_{i+1})_{H_i}}{T_{i+1} - T_i}\right] \Delta H_i, \quad (2)$$

where $M_i$ and $M_{i+1}$ are the experimental data of the magnetization at $T_i$ and $T_{i+1}$, respectively, under the same magnetic field.

Fig. 5 shows the thermal variation of the magnetic entropy change -$\Delta S_M$ under different ranges of magnetic field up to $\Delta H$ = 48 kOe. For each $\Delta H$, -$\Delta S_M$ reaches the maxima value at $T_C$, i.e., 279 K. The value of -$\Delta S_M$, which increases with increasing $\Delta H$, is 80.69mJ/cm$^3$ K and 133mJ/cm$^3$ K for the characteristic $\Delta H$ = 20 kOe and 48 kOe, respectively. A comparison between the peak value of -$\Delta S_M$ in SnCMn$_3$ and those in prototype MCE materials [3], such as Gd, Gd$_5$Si$_2$Ge$_2$, MnAs, DyCo$_2$, La$_{0.7}$Ca$_{0.3}$MnO$_3$ and LaFe$_{11.44}$Si$_{1.56}$, has been displayed in fig. 6. It is evident that -$\Delta S_M$ of SnCMn$_3$ is comparable to those of candidates for magnetic refrigerant materials plotted in this figure. Compared with GaCMn$_3$, the maxima -$\Delta S_M$ value of SnCMn$_3$ has a similar value, but happens near room temperature. Practically, the relative cooling power (RCP), which is a measure of how much heat can be transferred between the cold and hot sinks in an ideal refrigerant cycle, is also an important parameter for selecting potential substances for magnetic refrigerants. Generally, the RCP is defined as the product of the maximum magnetic entropy change $-\Delta S_M^{max}$ and full width at half maximum $\delta T_{FWHM}$ [3, 19],

$$RCP = -\Delta S_M^{max} \delta T_{FWHM} \qquad (3)$$

According to eq. (3), the estimated *RCP* of SnCMn$_3$ (~212.8mJ/cm$^3$, *ΔH* =48kOe) is less than those of Gd (5300mJ/cm$^3$, *ΔH* =50kOe) [3] and Gd$_5$Si$_2$Ge$_2$ (3360mJ/cm$^3$, *ΔH* =50kOe) [3]. Despite this, SnCMn$_3$ could be a candidate for magnetic refrigerant materials in the relative temperature region due to its inexpensive and innoxious raw materials, as well as considerable small thermal hysteresis.

Intuitively, as indicated by eqs. (1) and (2), the large -$\Delta S_M$ in SnCMn$_3$ can be attributed to the sharp change of magnetization in the vicinity of $T_C$ (fig. 2). For a better understanding of the origin of large MCE, the normal Hall effect and Seebeck coefficient of SnCMn$_3$ have been measured as a function of temperature. The $\alpha(T)$ (fig.7) is proportional to temperature at low temperatures, indicating the main contribution to $\alpha(T)$ comes from the thermal diffusion of carriers. Thus, the value of $\alpha(T)$ can reflect the type of carriers [20], namely, the negative $\alpha$ value means electron-type carriers in SnCMn$_3$. Another remarkable feature of $\alpha(T)$ is the remarkable change occurring at $T_C$. In a Boltzmann picture, the Seebeck coefficient $\alpha$ can be described as [21],

$$\alpha = -\frac{\pi^2 \kappa_B^2}{3e} T \left(\frac{\partial \ln N(E)}{\partial E}\right)_{E_F} \qquad (4)$$

where $N(E_F)$ is the electronic density of state (DOS) at Fermi energy $E_F$ and $k_B$ the Boltzmann constant. Eq. (4) indicates that any change in the Seebeck coefficient is a direct consequence of the modification in the DOS near $E_F$. Fig. 8 displays the temperature dependence of normal Hall coefficient, $R_H(T)$, measured at a magnetic field of 10 kOe. $R_H$ is negative in the entire temperature range investigated, indicating the dominant carriers are electron-type. As temperature increases, a sudden change of $R_H(T)$ occurs at $T_C$. The Hall carrier density $n_H$ can be reasonably estimated with the formula $n_H = 1/|eR_H|$ [20], where $e$ is elementary electric charge. As shown in inset of fig. 8, around $T_C$, $n_H$ in high-temperature PM phase is increased by three times in comparison with that in low-temperature FIM phase. The calculated value of $n_H$ at room temperature for SnCMn$_3$ (1.03×10$^{22}$ cm$^{-3}$) is a little less than that for common metal like Cu (~8.5×10$^{22}$ cm$^{-3}$), in line with the metal-like character of electrical conductivity. Consequently, the abrupt changes of $\alpha(T)$ and $R_H(T)$ in SnCMn$_3$ may indicate evidently a sudden reconstruction of the electronic structure, e.g., DOS near $E_F$. Such a change of DOS could not arise from the volume collapse at phase transition which usually leads to a reduction of DOS (namely carrier density) [22].

As discussed above, the large MCE in SnCMn$_3$, accompanied by abrupt changes of magnetization, resistivity and lattice, originates from the reconstruction of electronic structure at FIM-PM transition temperature. Such kind of behavior seems to be universal in Mn-based antiperovskite compounds, since sharp magnetic/structural phase transitions have been proved to be characteristic of them [15]. In addition, the types of phase transition and the corresponding temperatures can be tuned by varying chemical composition [9-12, 15]. For example, the properties of GaCMn$_3$ can be reproduced by alloying 50% SnCMn$_3$ with 50% ZnCMn$_3$ [15]. Therefore, it is possible to get a higher value of RCP by increasing the MCE or broadening its temperature span via chemical doping. In this regard, this type of materials can be considered as an alternative for pursuing new materials with large MCE at various temperatures. On the other hand, the mechanism of magnetic orders in Mn-based antiperovskite compounds is far away from clear. Initial work suggests that Ruderman-Kittel-Kasuya-Yosida (RKKY), which connects the itinerant electrons and localized magnetic moments, may work in these materials [15]. Alternatively, a simple mechanism of direct exchange interaction between Mn sites has been employed to explain the experimental results [11, 14]. Theoretically, a well-dispersed $d$ orbital has been found [13], suggesting an itinerant mechanism for magnetic orders, consistent with the result of Mössbauer spectroscopy [23]. In this scenario, however, it is a challenge to understand the first-order magnetic transition with abrupt change of lattice [23, 24]. In this context, more researches on Mn-based antiperovskite compounds will give a deeper insight into the nature of magnetic orders and competition between them, and vice versa.

In summary, we report the large magnetic entropy near room temperature in antipervoskite metallic compound SnCMn$_3$. The large value of $-\Delta S_M$ is comparable to those observed in contemporary magnetic refrigerant materials. The large magnetic entropy change is suggested by thermal transport measurements to be associated with the reconstruction of the Fermi surface in the vicinity of FIM-PM transformation. Along with the advantages of the raw materials, the discovery of MCE in SnCMn$_3$ may provide a new field for seeking after good candidates for magnetic refrigeration at various temperatures.

***

This work was supported by the National Key Basic Research under contract No. 2007CB925002, and the National Nature Science Foundation of China under contract No.50701042, No.10774146

**References**

[1] Tishin A. M., *J. Magn. Magn. Mater.*, **316** (2007) 351.

[2] Phan M. H. and Yu S. C., *J. Magn. Magn. Mater.*, **308** (2007) 325.

[3] Gschneidner Jr K. A., Pecharsky V. K. and Tsokol A. O., *Rep. Prog. Phys.*, **68** (2005) 1479.

[4] He T., Huang Q., Ramirez A.P., Wang Y., Regan K. A., Rogado N., Hayward M. A., Haas M. K., Slusky J. S., Inumara K., Zandbergen H. W., Ong N. P., and Cava R. J., *Nature (London).*, **411** (2001) 54.

[5] Uehara M., Yamazaki T., Kôri T., Kashida T., Kimishima Y., and Hase I., *J. Phys. Soc. Jpn.*, **76** (2007) 034714.

[6] Kamishima K., Goto T., Nakagawa H., Miura N., Ohashi M., Mori N., Sasaki T., and Kanomata T., *Phys. Rev. B*, **63** (2000) 024426.

[7] Tohei T., Wada H. and Kanomata T., *J. Appl. Phys.*, **94** (2003) 1800.

[8] Yu M. H., Lewis L. H., and Moodenbaugh A. R., *J. Appl. Phys.*, **93** (2003) 10128.

[9] Takenaka K. and Takagi H., *Appl. Phys. Lett.*, **87** (2005) 261902.

[10] Takenaka K., Asano K., Misawa M., and Takagi H., *Appl. Phys. Lett.*, **92** (2008) 011927.

[11] Lewis L. H., Yoder D., Moodenbaugh A. R., Fischer D. A. and Yu M.H., *J. Phys.: Condens. Matter*, **18** (2006) 1677.

[12] Tohei T., Wada H. and Kanomata T., *J. Magn. Magn. Mater.*, **272-276** (2004) e585.

[13] Motizuki K. and Nagai H., *J. Phys. C: Solid State Phys.*, **21** (1988) 5251.

[14] Li Y. B., Li W. F., Feng W. J., Zhang Y. Q. and Zhang Z. D., *Phys. Rev. B*, **72** (2005) 024411.

[15] Fruchart D. and Bertaut E. F., *J. Phys. Soc. Jpn.*, **44** (1978) 781.

[16] Kaneko T., Kanomata T., and Shirakawa K., *J. Phys. Soc. Jpn.*, **56** (1987) 4047.

[17] Provenzano V., Shapiro A. J., Shull R. D., *Nature.*, **429** (2004) 853

[18] Mohapatra N., Iyer K. K., and Sampathkumaran E. V., E*ur. Phys. J. B*, **63** (2008) 451

[19] Gschneidner JR K. A., Pecharsky V. K. and Pecharsky A. O., et al, *Mater. Sci. Forum*, **315** (1999) 69-76

[20] Siebold Th. and Ziemann P., *Phys. Rev. B*, **51** (1995) 6328.

[21] Behnia K., Jaccard D. and Flouquet J., *J. Phys.: Condens. Matter*, **16** (2004) 5187.

[22] Park M.S., Giim J.S., Park S.H., Lee Y.W., Lee S.I. and Choi E.J., *Supercond. Sci. Technol*, **17** (2004) 274.

[23] Grandjean F. and Gérard A., *J. Phys. F: Metal Phys.*, **6** (1976) 451.

[24] Kim I. G., Jin Y. J. and Lee J. I., Freeman A. *J., Phys. Rev. B*, **67** (2003) 060407(R).

**Figure captions**

Fig. 1: X-ray diffraction pattern (solid curve) and Rietveld refinement result (crosses) for SnCMn$_3$ at room temperature. The vertical lines show the Bragg peak positions as denoted by the index (*hkl*). The difference between the data and the calculation is shown at the bottom. The goodness parameters of refinement are indicated.

Fig. 2: Temperature dependence of magnetization *M-T* under ZFC and FC modes at *H* = 100Oe and inverse susceptibility as a function of temperature. The FC *M-T* was measured on both warming (FCW) and cooling (FCC).

Fig. 3: Resistivity measured at zero field as a function of temperature. Inset shows the resistivity at zero field in cooling and warming cycles in the vicinity of $T_C$.

Fig. 4: Magnetization isotherms of SnCMn$_3$ at various temperatures from 272 K to 288 K in the presence of external magnetic fields up to 48 kOe.

Fig. 5: Magnetic entropy change of SnCMn$_3$ for the field changing from 0-5 kOe to 0-48 kOe.

Fig. 6: The maximal $-\Delta S_M$ values of SnCMn$_3$ and several prototype magnetocaloric materials (derived from reference 3) as a function of the external magnetic field. The temperatures to which the maximal $-\Delta S_M$ correspond are marked in the brackets after the compounds.

Fig. 7: Temperature dependence of Seebeck coefficient *α*(*T*).

Fig. 8: Normal Hall coefficient as a function of temperature, $R_H$(*T*). Inset, temperature dependent

Hall carrier density $n_H(T)$ near $T_C$. Solid lines are guidance for eyes.

**Figures:**

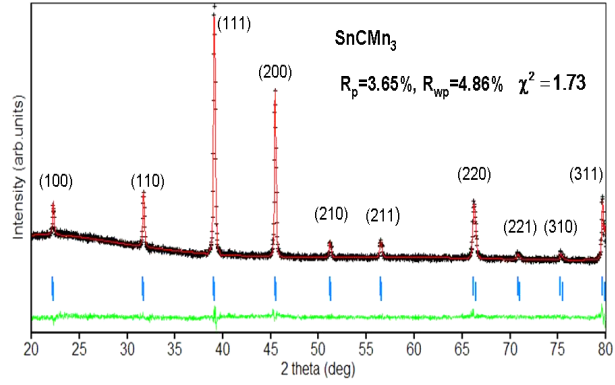

**Fig.1** Wang et al.,

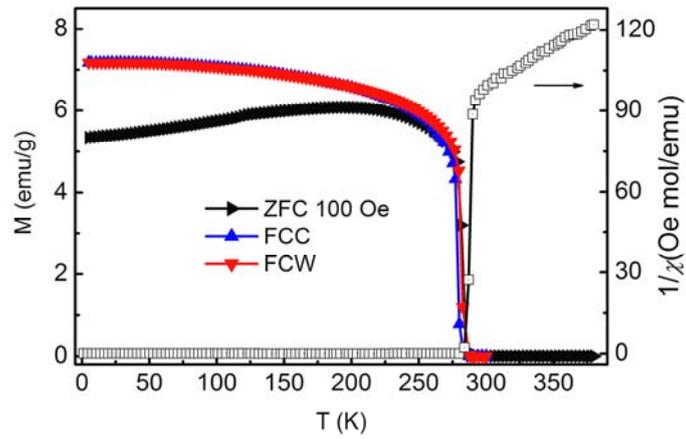

**Fig.2** Wang et al.,

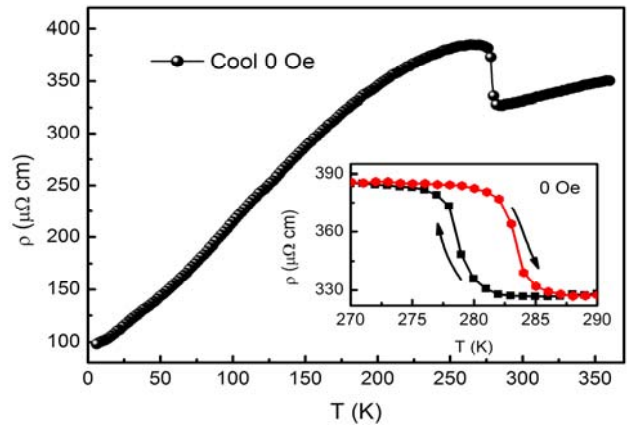

**Fig.3** Wang et al.,

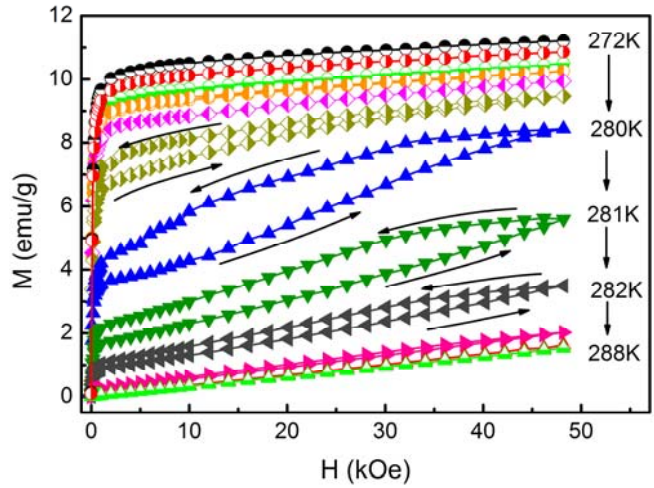

**Fig.4** Wang et al.,

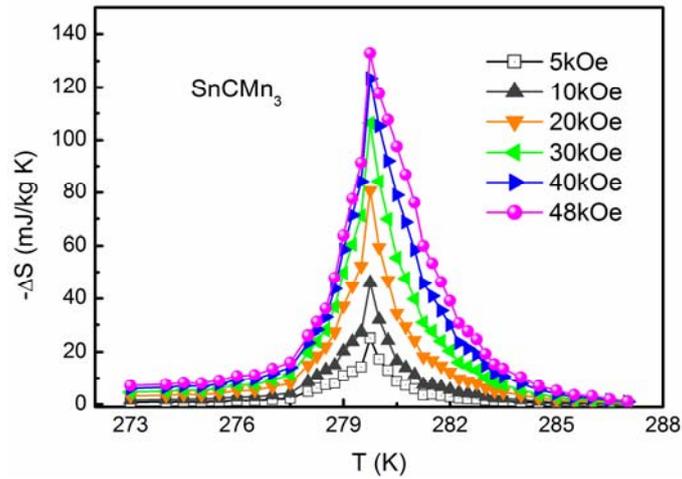

**Fig.5** Wang et al.,

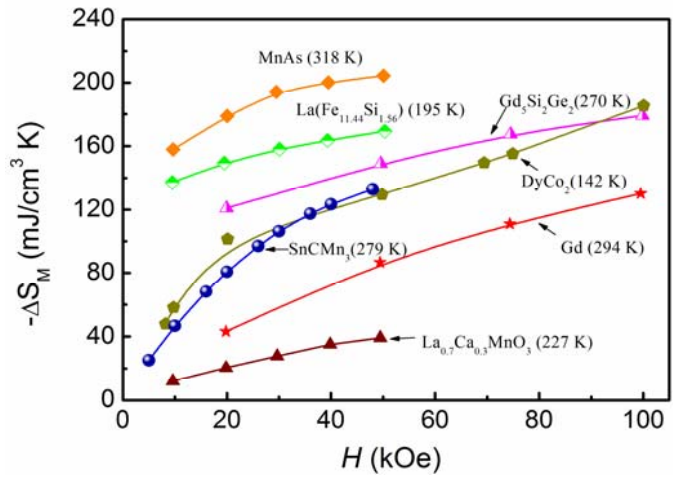

**Fig.6** Wang et al.,

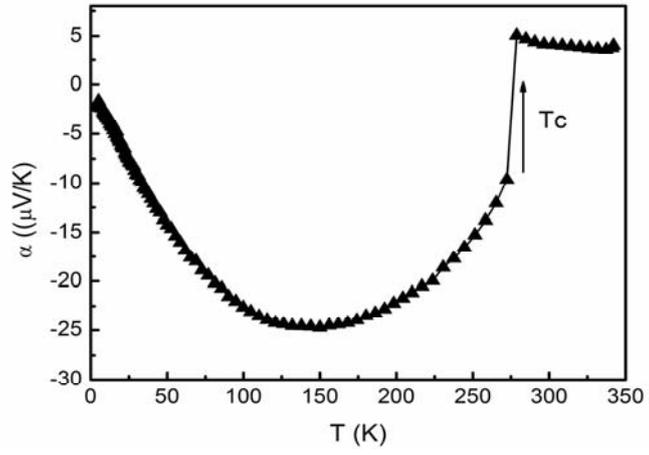

**Fig.7** Wang et al.,

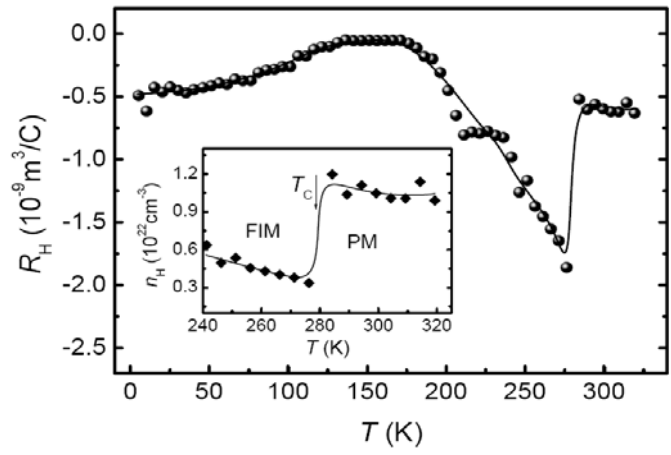

**Fig.8** Wang et al.,